\documentclass[conference, anonymous]{IEEEtran}
\IEEEoverridecommandlockouts
\usepackage{cite}
\usepackage{amsmath,amssymb,amsfonts}

\usepackage{hyperref}
\usepackage{graphicx}
\usepackage{fontawesome5}
\usepackage{tablefootnote}
\usepackage{textcomp}
\usepackage{xcolor}

\usepackage{amsmath,amssymb,amsfonts}
\usepackage{graphicx}
\usepackage{textcomp}
\usepackage[most]{tcolorbox} 
\def\BibTeX{{\rm B\kern-.05em{\sc i\kern-.025em b}\kern-.08em
    T\kern-.1667em\lower.7ex\hbox{E}\kern-.125emX}}
\usepackage[most]{tcolorbox} 
\usepackage{xspace} 
\usepackage[normalem]{ulem}  
\useunder{\uline}{\ul}{}  
\usepackage{booktabs} 
\usepackage{multirow} 
\usepackage{subcaption} 
\usepackage[noend]{algpseudocode}
\usepackage{enumitem}
\usepackage{adjustbox}
\usepackage{wrapfig}
\usepackage[ampersand]{easylist}
\usepackage[most]{tcolorbox}

\usepackage{colortbl}
\usepackage{amssymb}

\usepackage{pifont}
\newcommand{\xmark}{\ding{55}}

\definecolor{main}{HTML}{5989cf}    
\definecolor{sub}{HTML}{cde4ff}     

\newtcolorbox{boxB}{
    fontupper = \bf\color{main}\footnotesize, 
    boxrule = 0.5pt,
    colframe = main,
    rounded corners,
    arc = 5pt   
}

\newtcolorbox{boxD}{
    fontupper = \small, 
    colback = sub, 
    colframe = main, 
    boxrule = 0pt, 
    toprule = 2pt, 
    bottomrule = 2pt 
}

\newtcolorbox{boxH}{
    fontupper = \small, 
    colback = sub, 
    colframe = main, 
    boxrule = 0pt, 
    leftrule = 6pt 
}

\newtcolorbox{boxG}{
    enhanced,
    boxrule = 0pt,
    colback = sub,
    borderline west = {1pt}{0pt}{main}, 
    borderline west = {0.75pt}{2pt}{main}, 
    borderline east = {1pt}{0pt}{main}, 
    borderline east = {0.75pt}{2pt}{main}
}    

\newtcolorbox{boxK}{
    fontupper = \small,
    sharpish corners, 
    boxrule = 0pt,
    toprule = 1.0pt, 
    enhanced,
    fuzzy shadow = {0pt}{-2pt}{-0.5pt}{0.5pt}{black!35} 
}

\begin{document}
\newcommand{\linebreakand}{%
  \end{@IEEEauthorhalign}
  \hfill\mbox{}\par
  \mbox{}\hfill\begin{@IEEEauthorhalign}
}

\definecolor{MidnightBlue}{HTML}{006895}

\newcommand{\projtodo}[1]{{\color{red}{\bf TODO:}#1}}
\newcommand{\JSgruCloneIII}{{$0.81$}\xspace}
\newcommand{\JStfCloneIII}{{$1.0$}\xspace}

\newcommand*\circled[1]{\tikz[baseline=(char.base)]{
            \node[shape=circle,draw,inner sep=0.5pt] (char) {#1};}}

\newboolean{showcomments}

\setboolean{showcomments}{true}

\ifthenelse{\boolean{showcomments}}
  {\newcommand{\nb}[2]{
    \fbox{\bfseries\sffamily\scriptsize#1}
    {\sf\small$\blacktriangleright$\textit{#2}$\blacktriangleleft$}
   }
   \newcommand{\cvsversion}{\emph{\scriptsize$-$Id: macro.tex,v 1.9 2005/12/09 22:38:33 giulio Exp $}}
  }
  {\newcommand{\nb}[2]{}
   \newcommand{\cvsversion}{}
  }

\newcommand\myparagraph[1]{\noindent\underline{\bf {#1}:}}
\newcommand\myparagraphnew[1]{\noindent{\bf {#1}:}}
\newcommand\emparagraph[1]{\noindent {\em {#1}:}}
\newcommand\budget[1]{{\color{red}\myparagraph{Budget}{#1} pages}}
\newcommand\KEVIN[1]{{\color{ForestGreen} \nb{KEVIN}{#1}}}

\newcommand\DIPIN[1]{{\color{red} \nb{Dipin}{#1}}}

\newcommand\DANIEL[1]{{\color{red} \nb{DANIEL}{#1}}}
\newcommand\DAVID[1]{{\color{blue} \nb{DAVID}{#1}}}
\newcommand\DENYS[1]{{\color{blue} \nb{DENYS}{#1}}}
\newcommand\ALEJO[1]{{\color{purple} \nb{ALEJO}{#1}}}
\newcommand\TODO[1]{{\color{red} \nb{TODO}{#1}}}

\newcommand{\cancel}[1]{{\leavevmode\color{RubineRed}{\sout{\xspace#1}}}}
\newcommand{\edit}[2]{{\leavevmode\color{RubineRed}{\sout{#1}}}{\color{blue}{\xspace#2}}}
\newcommand{\rewrite}[2]{{\leavevmode\color{RubineRed}{\sout{#1}}}{\color{Green}{\arrow\xspace#2}}}

\newcommand{\add}[1]{{\leavevmode\color{blue}{\xspace#1}}}
\newcommand{\addnew}[1]{{\leavevmode\color{Green}{\xspace#1}}}
\newcommand{\remove}[1]{{\leavevmode\color{red}{\xspace#1}}}

\newcommand\finding[1]{\vspace{0.25em}\noindent\textsf{\bf Finding {#1}.}}
\newcommand\fnumber[1]{{$\mathcal{F}_{#1}$}}
\newcommand\operator[2]{{\bf OP$_{#1}$: {\em {#2}} -- }}
\newcommand\opnumber[1]{{{\bf OP}$_{#1}$}}

\newcommand{\arrow}{{$\rightarrow$}\xspace}
\newcommand\inline[1]{{\lstinline{#1}}}


\newcommand{\boxme}[1]{{
\begin{tcolorbox}[enhanced,skin=enhancedmiddle,borderline={1mm}{0mm}{MidnightBlue}]
    \textbf{Take Aways: } #1 \end{tcolorbox} 
}}

\newcommand\fix[1]{{\color{blue} \nb{FIX THIS}{#1}}}
\newcommand\blue[1]{{\color{blue}{#1}}}
\newcommand{\here}{{\color{blue} \nb{***}{CONTINUE HERE}}}

\newcommand{\REF}{{\color{red} \textbf{[REFS]}}\xspace}
\newcommand{\xy}{{\color{red} \textbf{XY}}\xspace}
\newcommand\tops[1]{{\color{blue}{#1}}}
\newcommand\alert[1]{{\color{red}{#1}}}

\newcommand{\target}{\textit{target tool}\xspace}
\newcommand{\targets}{\textit{target tools}\xspace}
\newcommand{\behavior}{\textit{target behavior}\xspace}
\newcommand{\ie}{\textit{i.e.,}\xspace}
\newcommand{\eg}{\textit{e.g.,}\xspace}
\newcommand{\etc}{\textit{etc.}\xspace}
\newcommand{\etal}{\textit{et al.}\xspace}
\newcommand{\etals}{et al.'s\xspace}
\newcommand{\aka}{\textit{a.k.a.}\xspace}	


\newcommand{\ntp}{\textit{NtP}\xspace}

\newcommand{\dlse}{\textsl{\small DL4SE}\xspace}
\newcommand{\ct}{\textit{c\&t}\xspace}

\newcommand{\nlms}{NCMs\xspace}
\newcommand{\nlm}{NCM\xspace}

\newcommand{\llms}{LLMs\xspace}
\newcommand{\llm}{LLM\xspace}
\newcommand{\llmc}{LLMc\xspace}

\newcommand{\chatgpt}{ChatGPT\xspace}

\newcommand{\datainterI}{\textit{ProgramRepair}\xspace}
\newcommand{\datainterII}{\textit{SemanticPreserving}\xspace}
\newcommand{\datainterIII}{\textit{UnCommenting}\xspace}

\newcommand{\modelinterI}{\textit{NumberLayers}\xspace}
\newcommand{\modelinterII}{\textit{NumberUnits}\xspace}

\newcommand{\randomCut}{\textit{RandomCut}\xspace}
\newcommand{\withDocstring}{\textit{WithDocString}\xspace}
\newcommand{\fromDocstring}{\textit{FromDocString}\xspace}
\newcommand{\commitGen}{\textit{CommitGen}\xspace}
\newcommand{\summarizationGen}{\textit{SummarizationGen}\xspace}

\newcommand{\ASTerrors}{\textit{n\_ast\_errors}\xspace}
\newcommand{\ASTlevels}{\textit{n\_ast\_levels}\xspace}
\newcommand{\ASTnodes}{\textit{n\_ast\_nodes}\xspace}
\newcommand{\whitespaces}{\textit{n\_whitespaces}\xspace}
\newcommand{\complexity}{\textit{complexity}\xspace}
\newcommand{\nloc}{\textit{nloc}\xspace}
\newcommand{\tokenCount}{\textit{token\_count}\xspace}
\newcommand{\identifiers}{\textit{n\_identifiers}\xspace}
\newcommand{\commitID}{\textit{commit\_id}\xspace}
\newcommand{\funName}{\textit{fun\_name}\xspace}
\newcommand{\commitMessage}{\textit{commit\_message}\xspace}
\newcommand{\docstring}{\textit{docstring}\xspace}
\newcommand{\promptSize}{\textit{prompt\_size}\xspace}
\newcommand{\control}{\textit{control}\xspace}
\newcommand{\Ta}{$T_1$\xspace}
\newcommand{\Tb}{$T_2$\xspace}

\newcommand{\assoJS}{JS Dist.\xspace}
\newcommand{\assoPR}{Pearson\xspace}


\newcommand{\rfi}{$\mathcal{R}_1$\xspace}

\newcommand{\galeras}{\textit{Galeras}\xspace}
\newcommand{\dataset}{\textit{CodeSearchNet}\xspace}

\newcommand{\blocks}{\texttt{\small[blocks]}\xspace}
\newcommand{\tests}{\texttt{\small[tests]}\xspace}
\newcommand{\oop}{\texttt{\small[oop]}\xspace}
\newcommand{\declarations}{\texttt{\small[declarations]}\xspace}
\newcommand{\exceptions}{\texttt{\small[exceptions]}\xspace}
\newcommand{\datatype}{\texttt{\small[datatype]}\xspace}
\newcommand{\loops}{\texttt{\small[loops]}\xspace}
\newcommand{\operators}{\texttt{\small[operators]}\xspace}
\newcommand{\conditionals}{\texttt{\small[conditionals]}\xspace}
\newcommand{\extra}{\texttt{\small[extraTokens]}\xspace}

\newcommand{\gptI}{\textit{gpt-3 [125M]}\xspace}
\newcommand{\gptII}{\textit{gpt-3 [1.3B]}\xspace}
\newcommand{\gptIII}{\textit{gpt-3 [2.7B]}\xspace}

\newcommand{\codegenI}{\textit{codegen-nl [350M]}\xspace}
\newcommand{\codegenII}{\textit{codegen-nl [2B]}\xspace}

\newcommand{\multiI}{\textit{multi-lang [110M]}\xspace}
\newcommand{\multiII}{\textit{multi-lang [350M]}\xspace}
\newcommand{\multiIII}{\textit{multi-lang [2B]}\xspace}

\newcommand{\monoI}{\textit{mono-lang [110M]}\xspace}
\newcommand{\monoII}{\textit{mono-lang [1.5B]}\xspace}
\newcommand{\monoIII}{\textit{mono-lang [350M]}\xspace}
\newcommand{\monoIIII}{\textit{mono-lang [2B]}\xspace}

\newcommand{\nlgpt}{\textit{NL GPT-3}\xspace}
\newcommand{\nlcodegen}{\textit{NL Codegen}\xspace}
\newcommand{\monolang}{\textit{Mono-Language-Type}\xspace}
\newcommand{\multilang}{\textit{Multi-Language-Type}\xspace}
\newcommand{\COVwmd}{\texttt{\small(wmd)}\xspace} 
\newcommand{\COVloc}{\texttt{\small(LoC)}\xspace} 
\newcommand{\COVreturn}{\texttt{\small(\#returns)}\xspace} 
\newcommand{\COVloop}{\texttt{\small(\#loops)}\xspace} 
\newcommand{\COVcomparison}{\texttt{\small(\#comparisons)}\xspace} 
\newcommand{\COVtry}{\texttt{\small(\#tryCatches)}\xspace} 
\newcommand{\COVparenthesized}{\texttt{\small(\#parenthesized)}\xspace} 
\newcommand{\COVexpression}{\texttt{\small(\#expressions)}\xspace} 
\newcommand{\COVnumber}{\texttt{\small(\#numbers)}\xspace} 
\newcommand{\COVstring}{\texttt{\small(\#stringLiterals)}\xspace} 
\newcommand{\COVmathops}{\texttt{\small(\#mathOps)}\xspace} 
\newcommand{\COVvari}{\texttt{\small(\#variables)}\xspace} 
\newcommand{\COVmnestedblock}{\texttt{\small(\#maxNextedBlocks)}\xspace} 
\newcommand{\COVanonyclass}{\texttt{\small(\#anonyClasses)}\xspace} 
\newcommand{\COVinnerclass}{\texttt{\small(\#innerClasses)}\xspace} 
\newcommand{\COVlambdaexp}{\texttt{\small(\#lambdaExpressions)}\xspace} 

\newcommand{\COVunique}{\texttt{\small(\#uniqueWords)}\xspace} 
\newcommand{\COVlog}{\texttt{\small(\#logStatements)}\xspace} 
\newcommand{\COVmod}{\texttt{\small(\#modifiers)}\xspace}

\newcommand{\secref}[1]{Sec.~\ref{#1}\xspace}
\newcommand{\figref}[1]{Fig.~\ref{#1}\xspace}
\newcommand{\tabref}[1]{Table~\ref{#1}\xspace}
\newcommand{\Phase}{{\sc Phase}\xspace}
\newcommand{\Phases}{{\sc Phase's~}\xspace}

\newcommand{\emphquote}[1]{{\emph{`#1'}}\xspace}
\newcommand{\emphdblquote}[1]{{\emph{``#1''}}\xspace}

\newcommand{\emphbrack}[1]{\emph{[#1]}\xspace}
			
\newcommand{\subj}{\emphbrack{subject}}
\newcommand{\act}{\emphbrack{action}}
\newcommand{\obj}{\emphbrack{object}}
\newcommand{\prep}{\emphbrack{preposition}}
\newcommand{\objtwo}{\emphbrack{object2}}

\newcommand*\ciclednum[1]{\raisebox{.5pt}{\textcircled{\raisebox{-.9pt}
{#1}}}}

\bstctlcite{IEEEexample:BSTcontrol}

\title{
Benchmarking Causal Study to Interpret Large Language Models for Source Code
}


\author{\IEEEauthorblockN{Daniel Rodriguez-Cardenas,
David N.~Palacio, Dipin Khati, Henry Burke and
Denys Poshyvanyk}
\IEEEauthorblockA{Department of Computer Science,
William \& Mary\\
Williamsburg, VA\\
Email: dhrodriguezcar, danaderpalacio, dkhati, hqburke, dposhyvanyk\{@wm.edu\}}}

\maketitle


\begin{abstract}

One of the most common solutions adopted by software researchers to address code generation is by training Large Language Models (\llms) on massive amounts of source code. \llms are rooted in the concept of emergent capabilities in which machines statistically learn complex patterns from code data. Although a number of studies have shown that \llms have been effectively evaluated on popular accuracy metrics (\eg BLEU, CodeBleu), previous research has largely overlooked the role of Causal Inference as a fundamental component of the interpretability of \llms' performance. Existing benchmarks and datasets are meant to highlight the difference between the expected and the generated outcome, but do not take into account confounding variables (\eg lines of code, number of tokens, prompt size) that equally influence the accuracy metrics. The fact remains that, when dealing with generative software tasks by \llms, no benchmark is available to tell researchers how to quantify neither the causal effect of SE-based treatments nor the correlation of confounders to the model's performance. In an effort to bring statistical rigor to the evaluation of \llms, this paper introduces a benchmarking strategy named \galeras comprised of curated testbeds for three SE tasks (\ie code completion, code summarization, and commit generation) to help aid the interpretation of \llms' performance.  

We illustrate the insights of our benchmarking strategy by conducting a case study on the performance of \chatgpt under distinct prompt engineering methods. The results of the case study demonstrate the positive causal influence of prompt semantics on \chatgpt's generative performance by an \textit{average treatment effect} of $\approx 3\%$. Moreover, it was found that confounders such as prompt size are highly correlated with accuracy metrics ($\approx 0.412$). The end result of our case study is to showcase causal inference evaluations, \textit{in practice}, to reduce \textit{confounding bias}. By reducing the bias, we offer an interpretable solution for the accuracy metric under analysis.

\end{abstract}

\IEEEpeerreviewmaketitle


\section{Introduction}\label{sec:introduction}


Deep Learning for Software Engineering (\dlse) is an emerging research area in the field of software maintainability that entails a paradigm shift in the form by which machines statistically learn complex patterns from code data. To support actionable downstream SE tasks (\eg code completion, code summarization, or commit generation), ample evidence supports that \dlse approaches in the form of Language Models are able to generate code conditioned on a well-defined prompt \cite{watson_systematic_2021, zhao_survey_2023, zan_large_2023}. While essential, \dlse approaches have been reduced to a group of large and self-supervised neural architectures (\ie Large Language Models or simply LLMs) comprised of multiple self-attention layers that perform linear transformations to extract salient features from programming and natural language data. In particular, Large Language Models for Code (\llmc) have led to a renewed interest in the automation of software engineering tasks. Most of this automation is a generative process in which underlying code and natural language features interact with each other to auto-complete\cite{austin2021program, Hendrycks2021apps, chen_generation_2021,White.MSR.2015,Ciniselli.TSE,Ciniselli.MSR}, summarize\cite{Hussain2020DeepTL, leclair_ensemble_2021,Moran.SANER.2022}, review \cite{Mastropaolo.TSE, Mastropaolo.ICSE, Tufano.ICSE.2021, Tufano.ICSE.2022}, trace \cite{Moran.Traceability.ICSE} and translate code \cite{Nguyen:ICSE15}; generate test cases \cite{white_reassert_2020, Raychev2014CodeCW,Watson.Asserts.ICSE}, detect cone clones \cite{White.ASE2016,Tufano.MSR.2018} or fix bugs \cite{Tufano2019LearningBugFixes,zhou_devign_nodate,SANER.2019,Tufano.ICSE19.Changes,Tufano2019,ASE.2018,Zimin.Sequencer,CanWeFix}. In fact, \llmc have been deployed in large-scale solutions to provide code generative services. Tools such as \chatgpt and GitHub Copilot, which are based on the \textit{gpt} architecture, exhibit good performance at the aforementioned tasks \cite{zhao_survey_2023}. 

Therefore, an increased interest has emerged in further evaluating these \llmc \cite{liu2023code,liu2023improving,xu_systematic_2022,chen_evaluating_2021} to standardize the quality assessment of the generated code. Unfortunately, the current evaluation process overly-relies on accuracy metrics leaving no consensus as to what other features or properties are impacting the code generation process. In other words, we require to control for factors that influence the performance of \llmc if our goal is to \textit{interpret} models' output. Few studies have sought to examine accuracy metrics from a causal perspective to interpret \llmc \cite{palacio_toward_2023}. Ergo, the problem remains that, when attempting to understand the prediction performance of \llmc, no benchmarks are available to articulate causal queries. 

Previous research has largely overlooked the role of causal inference in evaluating \llmc. In fact, existing benchmarks are not without flaws to detect \textit{confounding bias}, which refers to the statistical ability to control for variables that can influence models' performance beyond the SE treatments under study (\ie evaluating the best prompting method). That is, we study causation because we need to understand not only \textit{what} but also \textit{why} \llmc arrive at performance decisions. To overcome these challenges, we pose a code-based benchmarking strategy, named \galeras, to interpret \llmc concentrated on answering causal queries of interest. \galeras enables SE researchers to explain \llmc performance decisions from a curated set of code-based confounders, which are associated with a given SE treatment under study. \galeras is comprised of three parts: 1) seven testbeds for evaluating distinct SE downstream tasks free of sampling bias and data snooping, 2) a set of confounders to compute causal effects, and 3) a pipeline to curate data from open repositories. 


To illustrate how to exploit \galeras to interpret \llmc, we conducted a causal study to quantify the impact of confounding variables on \chatgpt's prediction performance to assess whether certain types of \textit{prompt engineering} methods are excelling at automating code completion tasks. Prompt engineering is associated with the emergent ability of \llms to learn from prompts (\ie in-context learning). This ability comprises a set of techniques that manipulates the structure of a \llm's input sequence to attain better and less computationally expensive outputs than applying other downstream methods such as fine-tuning \cite{liu2023improving}. We organize our study around two RQs that are fundamentally centered on the problem of \textit{prompt engineering} for code:

\begin{enumerate}[label=\textbf{RQ$_{\arabic*}$}, ref=\textbf{RQ$_{\arabic*}$}, wide, labelindent=5pt]\setlength{\itemsep}{0.2em}
      \item \label{rq:exploratory} { \textbf{Exploratory Analysis:} \textit{How different is the distribution of tokens between the generated and ground-truth code?}} 
      \item \label{rq:causal} {\textbf{Causal Analysis:} \textit{To what extent the type of Prompt Engineering is influencing the code completion performance?}}
\end{enumerate}

The achieved results show that prompt engineering methods indeed causally impact the accuracy of the model by an \textit{Average Treatment of Effect} (ATE) of 3\% between the semantics of the prompt and the accuracy metric. Hence, choosing an adequate prompting strategy can positively influence the code completion performance of \chatgpt. To summarize, our key contributions are: 1) A filtered testbed with non-contaminated code snippets for \llmc benchmarking; 2) a set of (confounding) features (\eg Cyclo Complexity, \# of AST levels)  included in the testbed; 3) a pipeline to generate new testbeds for a given SE task; and 4) a causal inference benchmarking to interpret \llmc.


\section{Related Work}
\label{sec:related}


Considerable research attention has been devoted to data collection and benchmarking for \llmc. Tab.\ref{tab:benchmark-comparison} showcases eight qualitative properties that we use to compare three state-of-art benchmarks (\ie \textit{CodeXGLUE}, \textit{IdBench}, and \textit{MultiPL-E}) with \galeras. Firstly, Husain \etal introduced \textit{CodeSearchNet} for code retrieval automation \cite{husain2019codesearchnet}. Their datasets have been mostly employed to pre-train \llms rather than benchmarking software tasks. Later, researchers at Microsoft extended \textit{CodeSearchNet} and amalgamated 12 SE-related datasets for other relevant downstream tasks (\eg clone detection, refinement, translation) \cite{lu_codexglue_2021}. These datasets and benchmarks are known as \textit{CodeXGLUE}, which partially support some accuracy and distance metrics. Secondly, Wainakh \etal proposed \textit{IdBench} to evaluate generated identifiers by measuring similarity distances of semantic representations \cite{wainakh_evaluating_2019}. Finally, Chen \etal notably posed \textit{HumanEval} to validate the functional correctness of generated code \cite{chen_evaluating_2021}. Cassano \etal  amplified \textit{HumanEval} to create \textit{MultiPL-E} for code translation \cite{cassano_multipl-e_2022}. Although these three benchmarks have been successfully employed for evaluating \llmc, these benchmarking strategies were not conceived to address the \textit{interpretation} of models' outputs. 

\begin{table}[t]
\centering
\caption{SOTA Benchmark qualitative properties comparison.}
\label{tab:benchmark-comparison}

\scalebox{0.75}{%
\setlength{\tabcolsep}{5pt} 

\begin{tabular}{ccccccc}
\hline
\multicolumn{2}{c}{\textbf{}} & \multicolumn{1}{l}{} & \multicolumn{4}{c}{\textit{\textbf{Benchmarks}}} \\ \cline{4-7} 
\multicolumn{2}{c}{\textbf{Qualitative Properties}} & \multicolumn{1}{l}{} & \textbf{CodeXGLUE} & \textbf{IdBench} & \textbf{MultiPL-E} & \textbf{Galeras} \\ \hline
 & Clone detection &  & \checkmark & \xmark & \xmark & \xmark \\
 & Defect detection &  & \checkmark & \xmark & \xmark & \xmark \\
 & Type Inferring &  & \xmark & \checkmark & \xmark & \xmark \\
 & Summarization &  & \xmark & \xmark & \xmark & \checkmark \\
 & Code generation &  & \xmark & \xmark & \xmark & \checkmark \\
 & Commit generation &  & \xmark & \xmark & \xmark & \checkmark \\
 & Repair &  & \checkmark & \xmark & \xmark & \xmark \\
 & Translation &  & \checkmark & \xmark & \checkmark & \xmark \\
\multirow{-9}{*}{\textit{\begin{tabular}[c]{@{}c@{}}Software\\ Tasks\end{tabular}}} & Search &  & \checkmark & \xmark & \xmark & \xmark \\ \hline
 & code-code &  & \checkmark & \xmark & \xmark & \checkmark \\
 & code-text &  & \checkmark & \checkmark & \xmark & \checkmark \\
\multirow{-3}{*}{\textit{I/O}} & text-code &  & \checkmark & \xmark & \checkmark & \xmark \\ \hline
 & Identifiers &  & \xmark & \checkmark & \xmark & \xmark \\
 & Code line &  & \checkmark & \xmark & \xmark & \xmark \\
 & Method &  & \checkmark & \xmark & \checkmark & \checkmark \\
\multirow{-4}{*}{\textit{\begin{tabular}[c]{@{}c@{}}Output\\ Granularity\end{tabular}}} & Files &  & \checkmark & \xmark & \xmark & \xmark \\ \hline
 & Words &  & \xmark & \checkmark & \xmark & \xmark \\
 & Tokens &  & \checkmark & \xmark & \xmark & \checkmark \\
 & Snippets &  & \checkmark & \xmark & \checkmark & \checkmark \\
\multirow{-4}{*}{\textit{\begin{tabular}[c]{@{}c@{}}Type of\\ Datum\end{tabular}}} & Prompts &  & \xmark & \xmark & \xmark & \cellcolor[HTML]{EFEFEF}\checkmark \\ \hline
 & Size &  & 416K & 500 answers & 164 problems & 227K \\
\multirow{-2}{*}{\textit{Dimension}} & Languages &  & $\approx 12$ & 3 & 19 & 1 \\ \hline
 & BLEU &  & \checkmark & \checkmark & \xmark & \checkmark \\
 & CodeBLEU &  & \checkmark & \xmark & \xmark & \checkmark \\
 & Cloze testing &  & \checkmark & \xmark & \xmark & \xmark \\
 & Levenshtein &  & \xmark & \checkmark & \xmark & \checkmark \\
 & Accuracy &  & \checkmark & \xmark & \xmark & \xmark \\
\multirow{-6}{*}{\textit{\begin{tabular}[c]{@{}c@{}}Supported\\ Metrics\end{tabular}}} & Causal Effect &  & \xmark & \xmark & \xmark & \cellcolor[HTML]{EFEFEF}\checkmark \\ \hline
\textit{Prompt}\cite{liu2023improving} & Single-step &  & \xmark & \xmark & \checkmark & \checkmark \\
\textit{Engineering} & Multiple-step &  & \xmark & \xmark & \xmark & \cellcolor[HTML]{EFEFEF}\checkmark \\ \hline
 & Confounders &  & \xmark & \xmark & \xmark & \cellcolor[HTML]{EFEFEF}\checkmark tab.\ref{tab:statistical-description}\\
\multirow{-2}{*}{\textit{\begin{tabular}[c]{@{}c@{}}Causal\\ Evaluation\end{tabular}}} & Inference &  & \xmark & \xmark & \xmark & \cellcolor[HTML]{EFEFEF}\checkmark \\ \hline
\end{tabular}
\vspace{-3.90cm}

}
{
Shadowed cells indicate \galeras only.
}
\end{table}

As \llmc are quickly evolving due to data and hyperparameter augmentation, current models (\eg \chatgpt, AlfaCode, Copilot) could have been trained on samples already used for evaluation (\aka data snooping) and datasets such as \textit{BigQuery} \cite{bigquery}, \textit{BigPython}\cite{nijkamp2023codegen}, and the \textit{Pile} \cite{gao2020pile} have omitted the importance of interpreting \llmc' performance. \galeras, however, offers curated testbeds for enabling prompt engineering evaluation. This evaluation includes an interpretability analysis based on \textit{causal inference} in the form of Structural Causal Models (SCM). What is more, \galeras provides a pipeline to collect and access confounders and treatment data. Such data is plugged into the SCM to estimate the causal effects between treatments and outcomes. Estimating these casual effects promote statistical rigor in evaluating SE-based generative tasks. 

\section{ Testbed Curation Pipeline }
\label{sec:design}
This section considers our proposed pipeline to structure and collect required testbeds for the comparative causal evaluation of \llmc. \galeras is a benchmarking strategy that entails a software architecture solution for the curation process.

\label{sec:pipeline}
\begin{figure}[ht]
		\centering
		\includegraphics[width=0.5\textwidth]{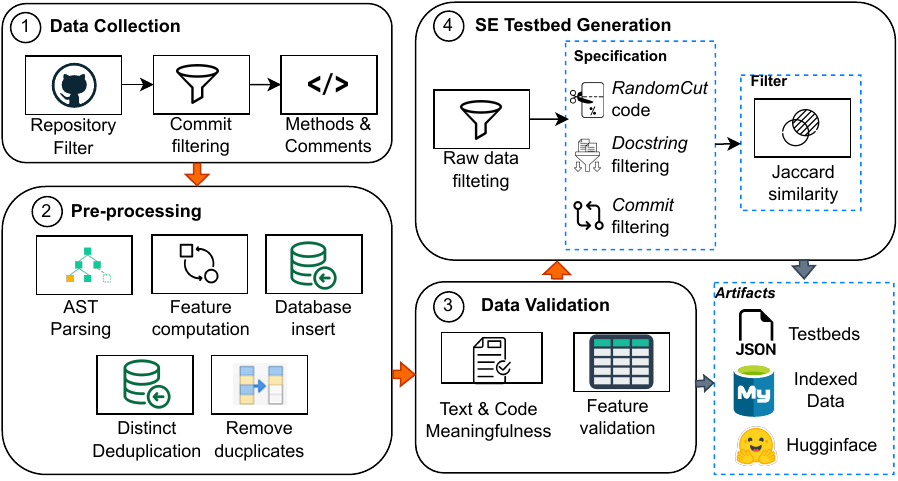}
		\caption{Testbed Curation Pipeline of \galeras}
        \vspace{-0.5cm}
        \label{fig:collection}
\end{figure}

\subsection{Structuring Testbed's Features}
\galeras testbeds are sets of Python methods that serve as evaluative data points. Each data point comprises four dimensions. The first dimension corresponds to snippets' identification,  which includes the \commitID (\ie commit hash), \textit{repository} name, \textit{path}, \textit{file\_name}, and \funName. The second dimension corresponds to snippets' documentation, which includes the \commitMessage and \docstring. The \docstring belongs to a JSON object that is extended to complementary natural language features such as \textit{n\_words, vocab\_size, language,} and \whitespaces. The third dimension corresponds to the snippet's syntactic information, which includes the actual code base, \ASTerrors, \ASTlevels, \ASTnodes,\textit{n\_words}, \textit{vocab\_size}, \tokenCount, and \whitespaces. Finally, the fourth dimension corresponds to canonical software metrics, which include \nloc, \complexity, and \identifiers.    





\subsection{Collecting Code Samples}

Figure \ref{fig:collection} describes a 4-step pipeline that \galeras uses to collect code samples. In the first step (Fig.~\ref{fig:collection}-\circled{1}), we filtered the most popular Python Github repositories using the following query: $language:Python$, $fork:false$, $size:>=30,000$,  $pushed:>2021-12-31$, $stars:>1,000$. From the last paper report of \chatgpt \cite{openai2023gpt4}, we assumed \chatgpt and other \llmc under analysis were not trained on commits from Jan 2, 2022 to Jan 1, 2023. Therefore, we claim that our testbeds help to avoid \textit{data snooping}, which is the misuse of data points to evaluate statistical hypotheses using training samples. Then, we collected a set of brand-new methods for each commit. This step resulted in $\approx338k$ data points. For each data point, we also collected its corresponding documentation without considering inline comments.

In the second step (Fig.~\ref{fig:collection}-\circled{2}), we engineered and preprocessed both code and documentation-related features from collected data points. Then we parsed the AST variables for our data points by employing the Tree-Sitter library. To guarantee efficient data management and once the previous features were engineered and extracted, we stored raw and preprocessed data points in a relational database. Next, we removed duplicated samples using a distinct query reducing the testbeds size to $\approx 227K$ data points for code (\textit{RawData} in tab.~\ref{tab:statistical-description}). Of these reduced data points, $\approx 77K$ contains a valid \docstring (\textit{RawDataDocstring} in tab.~\ref{tab:statistical-description}). A \docstring is valid when its text is larger than 3 words.

\begin{table*}[t]
\centering
\caption{Descriptive Analysis $[avg \pm std]$ of \galeras' Testbeds and Code Features.}
\vspace{-0.2cm}
\label{tab:statistical-description}

\scalebox{0.75}{%
\setlength{\tabcolsep}{5pt} 

\begin{tabular}{lclccclcccccc}
\hline
 &
  \multicolumn{1}{l}{} &
   &
  \multicolumn{3}{c}{\textbf{Confounders*}} &
   &
  \multicolumn{6}{c}{\textbf{Effect modifiers}} \\ \hline
\multicolumn{1}{c}{\textbf{Testbed}} &
  \textbf{Dedup} &
   &
  \textbf{n\_whitespaces} &
  \textbf{nloc} &
  \textbf{token\_counts} &
   &
  \textbf{n\_ast\_errors} &
  \textbf{ast\_levels} &
  \textbf{n\_ast\_nodes} &
  \textbf{complexity} &
  \textbf{token\_counts} &
  \textbf{n\_identifiers} \\ \hline
\textit{RawData} &
 277226 &
   &
  $259.23\pm902.22$ &
  $21.16\pm47.46$ &
  $137.38\pm262.59$ &
   &
  $0.09\pm0.42$ &
  $11.85\pm3.5$ &
  $221.91\pm438.23$ &
  $3.25\pm6.98$ &
  $137.38\pm262.59$ &
  $17.94\pm16.45$ \\
\textit{RawDataDocstring} &
  57045 &
   &
  $206.98\pm453.06$ &
  $18.89\pm30.98$ &
  $112.50\pm183.78$ &
   &
  $0.10\pm0.73$ &
  $11.57\pm3.45$ &
  $184.53\pm436.76$ &
  $3.42\pm6.61$ &
  $112.50\pm183.78$ &
  $15.96\pm14.48$ \\ \hline
\textit{RandomCut} &
  2931 &
   &
  $229.24\pm479.38$ &
  $18.27\pm26.98$ &
  $126.54\pm177.19$ &
   &
  $0.10\pm0.30$ &
  $12.25\pm3.06$ &
  $207.55\pm259.46$ &
  $3.16\pm6.09$ &
  $126.54\pm177.19$ &
  $17.70\pm13.42$ \\
\textit{WithDocstring} &
  2926 &
   &
  $208.48\pm414.67$ &
  $18.08\pm20.65$ &
  $111.98\pm122.27$ &
   &
  $0.08\pm0.58$ &
  $12.22\pm3.10$ &
  $188.99\pm400.74$ &
  $3.78\pm4.33$ &
  $111.98\pm122.27$ &
  $16.61\pm11.50$ \\ \hline
\textit{FromDocsting} &
  2937 &
   &
  $167.96\pm244.56$ &
  $16.68\pm20.91$ &
  $100.13\pm118.36$ &
   &
  $0.10\pm0.59$ &
  $11.38\pm3.44$ &
  $156.39\pm180.71$ &
  $3.48\pm4.48$ &
  $100.13\pm118.36$ &
  $14.62\pm11.94$ \\
\textit{CommitGen} &
  2919 &
   &
  $179.62\pm363.64$ &
  $16.75\pm21.37$ &
  $101.66\pm128.65$ &
   &
  $0.09\pm0.36$ &
  $11.51\pm3.36$ &
  $160.30\pm201.11$ &
  $3.28\pm4.93$ &
  $101.66\pm128.65$ &
  $15.07\pm11.90$ \\
\textit{SummarizationGen} &
  2924 &
   &
  $212.08\pm415.90$ &
  $18.66\pm21.63$ &
  $114.38\pm128.94$ &
   &
  $0.07\pm0.32$ &
  $12.22\pm3.16$ &
  $197.05\pm532.69$ &
  $3.85\pm5.36$ &
  $114.38\pm128.94$ &
  $16.71\pm12.16$ \\ \hline
\end{tabular}

\vspace{-0.45cm}

}
{ *The confounder \promptSize was omitted due to its treatment dependency. We measured its correlations in Tab.~\ref{tab:correlation}  \par}
\end{table*} 

In the third step \ref{fig:collection}-\circled{3}), we manually validated $960$ out of $\approx 227K$ data points. These validated data points were randomly selected from \textit{RawData} and \textit{RawDataDocstring}.  The remaining data points were automatically validated. Our validation process ensures the date of each pushed commit is within the range of dates stated in the original query. We also validated that the methods attached to each commit were indeed updated within the same range of dates. In addition, we validated the meaningfulness of the \docstring and \commitMessage by inspecting the consistency of the natural language descriptions with the actual code implementation, removing $\approx1.9\%$ \textit{RawDataDocstring} obtaining $\approx57K$ datapoints (tab. \ref{tab:statistical-description}). Lastly, \complexity was validated using the Codalyze plugin in Visual Studio Code. For the sake of simplicity, we omit explaining all considered fine-grained validation steps in this paper. However, the reader can consult our online appendix for more information\cite{daniel23}. 
 
In the final step (Fig.\ref{fig:collection}-\circled{4}), we sampled $3k$ data points from \textit{RawData} testbed to build five additional testbeds, each one for a specific SE task. \galeras comprises  \randomCut,  \withDocstring and \fromDocstring for \textit{code completion}; \commitGen for \textit{code generation}; and \summarizationGen for \textit{code summarization}. These additional testbeds are described in Tab.~\ref{tab:statistical-description}. To build \randomCut, we chose data points with more than $10$ tokens or $100$ characters. Next, the data point is randomly cut after the method signature. To build \summarizationGen and \commitGen, we filtered the \textit{RawDataDocstring} data points with more than 10 words or 50 characters. After building the five testbeds, we removed duplicated snippets using the Jaccard similarity on preprocessed data points with BPE HuggingFace tokenizer.  Because the de-duplication between training and test sets was discarded (\ie no multiset threshold), we set $0.7$ as the similarity threshold for our testbeds \cite{Allamanis19,wang_neural_2019}. Table.~\ref{tab:dedupe} shows the SE Task associated with each curated testbed, the percentage rate of detected duplicates, and the final size.     

\begin{table}[t]
\centering
\caption{Jaccard Similarity de-duplication}
\label{tab:dedupe}

\scalebox{0.8}{%
\setlength{\tabcolsep}{5pt} 

\begin{tabular}{llcccc}
\hline
\multicolumn{1}{c}{\textbf{SE Task}} & \multicolumn{1}{c}{\textbf{Testbed}} & \multicolumn{1}{c}{\textbf{I/O}} & \textbf{Dupes} & \textbf{Dupe \%} & \textbf{size} \\ \hline
 & \randomCut & code $\Rightarrow$ code & 69 & 2.30\% & 2931 \\
 & \withDocstring & code-text $\Rightarrow$ code & 74 & 2.47\% & 2926 \\
\multirow{-3}{*}{\textbf{Code completion}} & \fromDocstring & text $\Rightarrow$ code & 63 & 2.10\% & 2937 \\
\textbf{Code generation} & \commitGen & code $\Rightarrow$ text & 81 & 2.70\% & 2919 \\
\textbf{Summarization} & \summarizationGen & code $\Rightarrow$ text & 76 & 2.53\% & 2924 \\ \hline
\end{tabular}
\vspace{-3.90cm}

}
\end{table}

\section{ Causal Analysis for Interpretable \llmc}
\label{sec:benchmark}

\begin{figure}[t]
		\centering
		\includegraphics[width=0.48\textwidth]{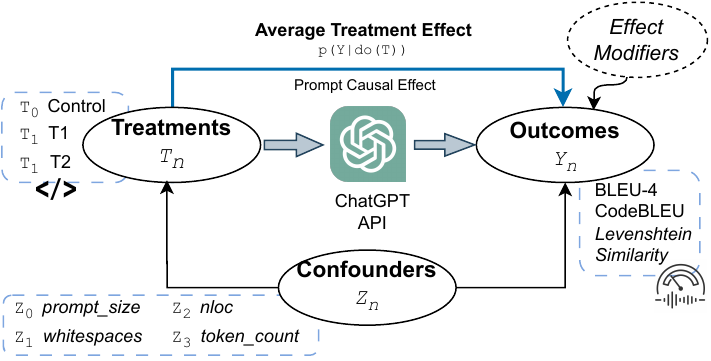}
		\caption{\galeras Structural Causal Model Benchmarking}
        \vspace{-0.5cm}
        \label{fig:case-study}
\end{figure}

\galeras is a causal benchmarking to compare the performance of \llmc against each other by controlling for \textit{confounding variables}, which are features of the source code that can influence the prediction performance of \llmc. Ideally, researchers can use \galeras to contextualize the outcomes of \llmc by presenting possible tailored treatment variables that explain the behavior of the model. \galeras' goal is to empower the research community to interpret typical performance metrics by stating the assumptions of the prediction problem in a \textit{Structural Causal Model} (SCM). The SCM comprises four random variables. The first variable is the \textit{treatments} $T$, which represents the input configuration prompts in our case study. The second variable is the \textit{potential outcomes} $Y$, which is the model prediction performance measured using distance metric (\eg BLEU, CodeBLEU, Levenshtein). The third variable is the \textit{confounders} $Z$, which represents variables affecting both $T$ and $Y$ (see Fig. \ref{fig:case-study}).  The last variable is the \textit{effect modifiers}, which is the features directly affecting outcomes $Y$. 

The purpose of the causal analysis is to eliminate \textit{spurious correlations} between the treatments $T$ and the outcomes $Y$ by controlling for confounding features $Z$. The elimination of the confounding features can be formally described with both an SCM and the $do$-operator introduced by Pearl \etal \cite{Pearl2009Causality}. We measure the \textit{Average Treatment Effect} (ATE) by approximating the conditional probability $p(Y|do(T))$ with statistical methods such as the propensity score matching, stratification, or IPW \cite{Pearl2009Causality,Sharma2021DoWhyAssumptions}. An in-depth analysis and explanation of causal inference methods are beyond the scope of this paper. 

\section{Causal Study: Interpretable Code Completion}
\label{sec:approach}

To demonstrate how to employ \galeras for causal analysis, in practice, we design a study in which we evaluate \chatgpt's performance for two prompt engineering methods $T_1$ and $T_2$ based on Liu \etal \cite{liu2023improving}. Prompt engineering is the activity of optimizing the input space of a given \llm in order to generate better outcomes without giving rise to expensive fine-tuning. The goal of our case study is to compare these two prompting methods after controlling for confounding features.  

\subsection{Evaluation Methodology}
The evaluation methodology of the case study is divided into three parts. The first part addresses the exploratory analysis of \galeras testbeds. We employed the BPE tokenizer to normalize the vocabulary of each treatment $T$ and outcome $Y$ sentence. The token count categorized by taxonomy is presented in Fig.\ref{fig:descriptive-analysis}. Tokens within each sentence were classified based on their taxonomy, \ie  \textit{`try'} and \textit{`catch'} are classified as \textit{exceptions} and  \textit{`if'} and \textit{`else'} as conditionals. Since the analysis focused solely on Python, keywords related to data types were classified as \textit{casting} tokens.

The second part canonically evaluates \chatgpt using our testbed \withDocstring. CodeBLEU was computed with a default parameter value of $0.25$. In addition, BLUE was computed with a 4-gram parameter. On the other hand, we computed the Levenshtein distance and similarity for a local evaluation (see Tab~.\ref{tab:correlation}-Performance Metrics). 

The third part estimates the causal effect of prompt engineering methods and \chatgpt performance. Figure \ref{fig:case-study} illustrates our Structural Causal Models for the prompt engineering case of \chatgpt. We use \galeras to compare the performance of two different treatments. The first treatment \Ta is one prompt, which contains a command (\eg \textit{Complete the following a Python code, return only code and complete method: `\{partial code\}'} ) followed by the actual input code to be completed. The second treatment \Tb comprises two prompts. The first one is a context prompt that entails both the \docstring and the incomplete cut code. The second one is a \textit{processing prompt} that contains sentences asking for removing comments and optimizing code (\eg Remember you have a Python function named `\{ \funName\}', the function starts with the following code `\{code\}'. The description for the function is: `\{ \docstring\}' ). We used the previous treatments against a \control group. The \control is a \textit{task prompt} that encompasses an action word or verb followed by the incomplete code input (\eg Complete the following python method: `\{partial code\}'). To evaluate whether treatments $T$ are impacting \chatgpt performance $Y$, we controlled for confounding features $Z$. Our confounders \promptSize, \whitespaces, \tokenCount, and \nloc were selected due to their high correlation ($[0.4-0.8]$) with the Levenstein distance in control and treatment groups. Although \ASTnodes has a high correlation with the Levenstein distance, we assumed that structural features are ignoring the treatments. Hence, AST-based features are effect modifiers. The potential outcomes $Y_2$,$Y_1$,$Y_0$ are observed under the treatments \Ta,\Tb,\control. Next, we approximate the \textit{Average Treatment Effect} $p(Y|do(T)$ using the SCM defined in Fig~.\ref{fig:case-study}.

\subsection{Results}
\ref{rq:exploratory} \textit{Exploratory Analysis.}  The purpose of the exploratory analysis is to expose and understand the testbeds' feature distribution grouped by prompt engineering methods $T$. Table \ref{tab:statistical-description} depicts the average and standard deviation for each code feature. We observed high variability in \whitespaces ($902.22$) and \tokenCount ($262.59$), which implies the method sizes are not homogeneous across the testbeds. While the descriptive analysis showcases high variability for all code features, our testbeds are a representative sub-sample of open repositories. For instance, the \complexity feature has an average value of $3.25$ suggesting that the code has a reasonable number of loops, conditionals, and operators. Therefore, our collected methods exhibit that our pipeline process guarantee data point diversity.   

We observed no significant differences in the counting of tokens among potential outcomes (including the \control) and the ground truth (see Fig.~\ref{fig:descriptive-analysis}-A). For instance, \control and \Tb on declarations (with a diff. around $550$ tokens) and loops (with a diff. around $600$ tokens) are relatively small. However, \Ta outcome exhibited high difference and excessive use of OOP, declarations, and loops with a diff. around $2.6k$, $2k$, and $1.5k$  tokens respectively. Figure \ref{fig:descriptive-analysis}-B showcases the token distribution for each testbed. We detected that the two prompt engineering methods were generating a similar amount of tokens (\ie green and red distributions) compared to the \control and ground truth. This suggests that sophisticated prompts tend to generate repetitive tokens. Figure \ref{fig:descriptive-analysis}-C depicts the Levenshtein similarity distance between the \chatgpt outputs, generated with both prompt engineering methods and the \control, and the ground truth. We can observe from the proportion curve that \Ta similarity performs the worst compared to the \control and \Tb.


\begin{boxK}
\ref{rq:exploratory} Exploratory Analysis: Grouped by taxonomy the ground truth does not repeat the same tokens as much as the treatments. The \Ta outcome seems to have notable intense use of keywords for OOP, declarations, and loops; \Tb obtains better performance with the highest similarity average of $0.43$
\end{boxK}

\begin{figure}[hb]
		\centering
		\includegraphics[width=0.5\textwidth]{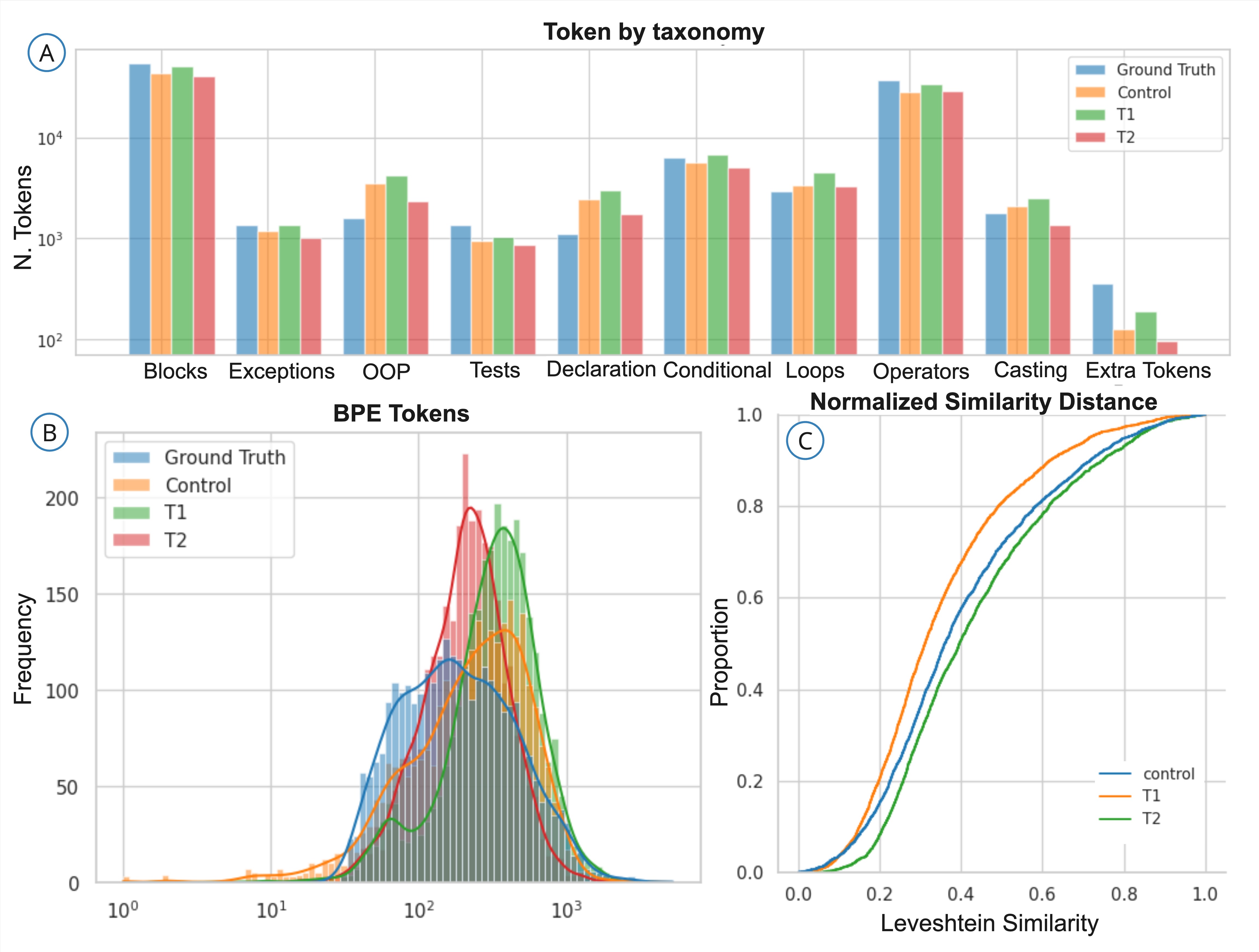}
		\caption{Descriptive Analysis: Top graph the token count for each testbed, bottom left the token frequency distribution, bottom right the similarity proportion score.}
        \vspace{-0.3cm}
        \label{fig:descriptive-analysis}
\end{figure}

\ref{rq:causal} \textit{Causal Analysis.} 

For two basic prompt engineering methods, code completion performance of \chatgpt is mainly affected by the following confounders: number of white spaces, lines of code, tokens in the outcome, and tokens in the prompt with a maximum correlation of $0.80$ with the Levenstein distance (see Tab.~\ref{tab:correlation}-Correlations). This suggests that after controlling for confounders, the \textit{Average Treatment Effect} (ATE) the prompt engineering method$_1$, represented by $T_1$, has a negative causal effect $p_1(Y|do(T)) = E[Y_1 - Y_0] \approx -5.1\%$ compared to a positive causal effect $p_2(Y|do(T)) = E[Y_2 - Y_0] \approx 3.3\%$ of method$_2$, represented by $T_2$ (see Tab.~\ref{tab:correlation}-Causal Effects). This indicates that method$_1$ is negatively affecting the Levenshtein similarity (\ie poor performance) across \withDocstring testbed, while method$_2$ is actually enhancing \chatgpt prediction performance. These results are consistent with the previous section in which we demonstrated that \Tb performs better than \Ta. After controlling for the confounding effect of the code features such as the prompt size and token counts, we can claim that the reason why \Tb is performing better than \Ta is \textit{purely} due to the information contained in the prompt. 

In order to validate the robustness of computed ATEs and proposed SCM, we refuted our effects using the following methods: \textit{Placebo, Random Common Cause (RCC)} and \textit{Subet} (see DoWhy refuters in \cite{Sharma2021DoWhyAssumptions}). We found that, for the ATEs computed with score matching, their corresponding refutation values are not stable. That is, the placebo value for $Y_1$ similarity is far from zero with $2.98$, while the RCC value differs by around $212$ in $Y_2$ distance. 

\begin{table}[ht]
\centering
\caption{Code Completion Testbed Results: 
Performance Metrics, Correlations, and Causal Effects.}
\vspace{-0.1cm}
\label{tab:correlation}

\scalebox{0.75}{%
\setlength{\tabcolsep}{5.5pt} 

\begin{tabular}{llcccccc}
\hline
\multicolumn{2}{l}{\textit{\textbf{Treatments}}} & \multicolumn{2}{c}{\textbf{Control}} & \multicolumn{2}{c}{\textbf{T1}} & \multicolumn{2}{c}{\textbf{T2}} \\ \hline
\multicolumn{8}{l}{\textbf{Performance Metrics}} \\ \hline
 \multirow{1}{*}{\textit{\textbf{Distance}}}& Bleu & \multicolumn{2}{c}{0.444} & \multicolumn{2}{c}{0.45} & \multicolumn{2}{c}{0.42} \\
 & CodeBleu & \multicolumn{2}{c}{0.441} & \multicolumn{2}{c}{0.438} & \multicolumn{2}{c}{0.469} \\
 \textit{\textbf{Similarity}}& Avg. Lev. & \multicolumn{2}{c}{0.40$\pm$0.20} & \multicolumn{2}{c}{0.35$\pm$0.18} & \multicolumn{2}{c}{0.43$\pm$0.20} \\ \hline
\multicolumn{2}{l}{\textbf{Correlations (vs Levenshtein)}} & \textbf{Dist.} & \textbf{Sim.\%} & \textbf{Dist.} & \textbf{Sim.\%} & \textbf{Dist.} & \textbf{Sim.\%} \\ \hline
\multirow{1}{*}{\textit{\textbf{Confounders}}} & \promptSize & 0.45 & 25.6\% & 0.40 & 41.2\% & 0.45 & 28.3\% \\
 & \whitespaces & 0.69 & 5.6\% & 0.62 & 20.7\% & \textbf{0.80} & \cellcolor[HTML]{FFFFFF}1.8\% \\
 & \tokenCount & 0.67 & 5.3\% & 0.59 & 24.8\% & 0.70 & 3.9\% \\
 & \nloc & 0.64 & 4.2\% & 0.57 & 20.7\% & 0.70 & 0.1\% \\
\multirow{1}{*}{\textit{\textbf{Effect Modifiers}}}  & \complexity & 0.43 & 4.3\% & 0.40 & 16.8\% & 0.47 & 0.9\% \\
 & \ASTnodes & 0.72 & 7.8\% & 0.62 & 29.4\% & 0.77 & 4.3\% \\
 & \ASTerrors & 0.02 & -2.4\% & 0.05 & 3.7\% & 0.18 & 2.3\% \\
 & \ASTlevels & 0.40 & 9.9\% & 0.31 & 30.4\% & 0.44 & 8.1\% \\ \hline
\multicolumn{8}{l}{\textbf{Causal Effects $(T\rightarrow Y)$}} \\ \hline
 \textit{\textbf{Score Matching}} & ATE & - & - & 104.02 & -3.7\% & -314.36 & 6.9\% \\
 & Placebo & - & - & -0.21 &   \textbf{298\%} & 0.02 & 0.1\% \\
 & RCC & - & - & 112.14 & -5.2\% & -102.71 & 3.3\% \\
 & Subset & - & - & 110.85 & -5.1\% & -101.6 & 3.3\% \\
\textit{\textbf{Stratification}} & ATE & - & - & 111.05 & -5.1\% & -101.73 & 3.3\% \\
 & Placebo & - & - & -0.17 & 0.04\% & 0.01 & {\ul0.05\%} \\
 & RCC & - & - & 111.17 & -5.1\% & -101.7 & 3.3\% \\
 & Subset & - & - & 110.95 & -5.2\% & -101.49 & 3.3\% \\
 \textit{\textbf{IPW}} & ATE & - & - & 111.05 & -5.1\% & -101.73 & 3.3\% \\
 & Placebo & - & - & -0.54 & -0.02\% & -1.30 & {\ul-0.07\%} \\
 & RCC & - & - & 111.04 & -5.1\% & -101.74 & 3.3\% \\
 & Subset & - & - & 111.12 & -5.1\% & -101.47 & 3.3\% \\ \hline
\end{tabular}
\vspace{-0.6cm}

}

{
 bold: highest correlation, underline: null effect. 
}
\end{table} 


\begin{boxK}
\ref{rq:causal} Causal Analysis: The prompt engineering method$_1$ (treatment \Ta) has a negative causal impact on the \chatgpt performance with an ATE estimation of $-5\%$. Conversely, the prompt engineering method$_2$ (treatment \Tb) has a subtle positive influence on the same performance with an ATE of $3\%$. This suggests that after controlling for prompt size, white spaces, \# of tokens, and nlocs; prompt engineering strategies are indeed affecting the quality of code completion.
\end{boxK}

\section{Conclusion \& Future Work}
\label{sec:conclusion}
This study used a qualitative technique to analyze the causal effect of SE-oriented treatments on the performance of \llmc. Such a technique is embedded into a benchmarking strategy named \galeras. Our benchmarking enables researchers to interpret \textit{why} a given \llmc is reporting a particular accuracy metric. We curated two raw Python testbeds: \textit{RawData} with only mined code and \textit{RawDataDocstring} with the corresponding documentation from GitHub. We also provide five SE Python testbeds for three SE tasks (\ie code completion, code summarization, and commit generation), we proposed a pipeline for collecting testbeds from git repositories. Finally, we conducted a rigorous evaluation of code completion with \chatgpt. Our causal study suggests that \chatgpt's performance is not only affected by the prompt size but also by the prompt semantics.  Future research will focus on determining whether other unmeasured confounders are affecting \llmc's prediction by augmenting the number of testbeds.

\section{Acknowledgement}\label{sec:acknowledgement}


This research has been supported in part by the NSF CCF-2311469, CNS-2132281, CCF-2007246, and CCF-1955853. We also acknowledge support from Cisco Systems. Any opinions, findings, and conclusions expressed herein are the authors’ and do not necessarily reflect those of the sponsors.

\bibliographystyle{IEEEtran}
\bibliography{IEEEabrv,references}
\end{document}